\documentclass[preprint,showpacs]{revtex4}
\usepackage{epsfig}
\usepackage{amsmath}

\begin{document}
\title{Spin glass behavior of the antiferromagnetic Heisenberg model on scale free network}
\author{Tasrief Surungan$^{1,3}$}
\email{tasrief@unhas.ac.id}
\author{Freddy P. Zen$^{2,3}$}
\email{fpzen@fi.itb.ac.id}
\author{Anthony G. Williams$^4$}
\email{anthony.williams@adelaide.edu.au}
\affiliation{$^1$Department of Physics, Hasanuddin University, Makassar 90245, Indonesia}
\affiliation{$^2$Department of Physics, Bandung Institute of Technology, Bandung 40132, Indonesia}
\affiliation{$^3$Indonesian Center for Theoretical and Mathematical Physics (ICTMP),
Bandung Institute of Technology, Bandung 40132, Indonesia}
\affiliation{$^4$Special Research Center for the Subatomic Structure of Matter (CSSM),
 The University of Adelaide, Adelaide, SA 5005, Australia }

\begin{abstract}
Randomness and frustration are considered to be the key 
ingredients for the existence of spin glass (SG)  phase. In a canonical
system, these ingredients are realized by the random mixture of 
 ferromagnetic (FM) and antiferromagnetic (AF)
couplings.  The study by  Bartolozzi {\it et al.}
[Phys. Rev. B{\bf 73}, 224419 (2006)] who observed
the presence of SG phase on the AF Ising model 
on scale free network (SFN) is stimulating.  It is a new type of SG
 system where randomness and frustration are not caused by the 
presence of  FM and AF couplings. To further elaborate this
type of system, here we study  Heisenberg  model 
on AF SFN and search for the SG phase.  The  canonical SG  Heisenberg model is not
observed in  $d$-dimensional regular lattices  for  ($d \leq 3$).
We can make an analogy for the connectivity density ($m$) of SFN with
the dimensionality of the regular lattice. It should be plausible
to find the critical value of $m$ for the existence of SG behaviour,
analogous to the lower critical dimension ($d_l$) for the canonical SG systems.
Here we study system with $m=2,3,4$ and $5$. We used Replica Exchange algorithm of  
Monte Carlo Method and calculated the SG order parameter.
We observed  SG phase for each value of $m$ and estimated its 
corersponding  critical temperature.
\end{abstract}

\maketitle

\section{Introduction}
Spin glass (SG) is one of the most complex systems in 
condensed matter physics and has been  intensively studied
in the last four decades\cite{Cannella, SK, Edwards,
Thirumalai, Kawamura, Wittmann}. 
It is a randomly frustrated magnetic system 
with frozen disordered spin orientation  at low temperatures.
This unusual configuration is regarded  as 
a temporal ordered phase \cite{Nishimori}, different from the spatially ordered 
phase found in regular magnets.  The complexity of the 
system is due to the presence of frustration and randomness, 
which are  the key ingredients for the existence of SG phase. 
Frustration is a state where  spins  can not 
find fixed orientations to fully satisfy all the interactions with their neighboring spins.
This can be  caused either by  the conflicting interaction
between  FM and AF couplings, or between among AF coupling due to the topological factors.  
Frustration alone can not lead a system to an SG phase, 
firmly exemplified by the fully frustrated AF planar spin 
systems which have spatially ordered phase at low temperatures\cite{Villain,Tasrief2004}.

Most SGs studied  are canonical system where 
both FM and  AF couplings exist. The examples of these are
Sherrington-Kirkpatrick model\cite{SK},  Edward-Anderson
model\cite{Edwards} and p-spin interaction model\cite{Thirumalai}.  
Bartolozzi {\it et al}. first reported SG behavior
of the Ising model with AF interaction on scale free network (SFN)\cite{Bartolozzi}.
This is a new  type of SG system  without random distribution of FM
and AF couplings.  The nodes of the network do not have homogeneous number of
links  and frustration is fully due to  the topological factor.  
The work has brought a new insight into the study of SGs, suggesting that
the irregular connectivity can also be one of the ingredients of SGs, different from 
the previous notion insisting the presence of random mixture  of FM  and AF interactions. 

An SFN consists of abundance of triangular units on which  spins 
 are frustrated if the couplings are AF.  While the frustration is caused by the
topological factor, the randomness is due to irregular connectivity.  
There are some vertices  having very large number of connections,
acting as the  hubs as in the internet connection.  In fact, the structure of 
the internet follows a scale free behavior.  The AF Ising model on random 
networks without a scale free behavior was also reported to exhibit SG phase
\cite{Herrero}.  We believe  that random connectivity can generally be an alternative 
for the usual randomness, together with frustration, as the  ingredients of SGs.

Here we study the Heisenberg SG model on SFN.  The prevalent controversy 
on existence of SG phase for the 3D canonical Heisenberg model  
 \cite{Kawamura} over the last three decades is one of the main motivations.
Most SGs  studied, such as CuMn, AgMn or CdMnTe, are
Heisenberg-like systems. 
%
Early numerical study by Coluzzi observed  SG phase transition  on 4D Heisenberg
 model.  Kawamura and Nishikawa pointed out the absence
of Heisenberg SG phase on D-dimensional space, for D $\leq 3$ \cite{Kawamura}. 
These works suggested that the lower critical dimension $d_l$ might be a fractional
number, $ 3 < d_l <4$. 

For a regular lattice, spatial dimension is related to the coordination number,
 i.e., the number of neighbors of each spin.  We can associate the coordination number 
with connectivity density $m$, which is the average number of links, of SFNs. 
Due to the presence of short-cut between  spins,
the notion of spatial dimension is not an  appropriate term for SFNs. Nonetheless,
 an SFN can still be associated with a  high dimensional regular lattice. 
Therefore,  we assume that there may also be a critical value of $m$, analogous to 
$d_l$ of regular lattice, for the random and SFNs.  
The paper is organized as follows: Section \ref{Model} describes
the models and method. The results are discussed in Section \ref{Results}.
Section \ref{Summary} is devouted for the summary and concluding remarks. 

\section{Model and Method of Simulation}\label{Model} 
The Heisenberg model on an SFN can be written with the following Hamiltonian,
\begin{equation}
H = -J \sum_{\langle ij \rangle } \vec s_i \cdot \vec s_j
\end{equation}
where for an AF system,  the coupling constant is set to be negative 
($J < 0$); and $\vec s_i$ are the Heisenberg spins residing  
 on the  nodes of the network.  The summation is performed over all 
directly connected neighbors. In an SFN, the number of neighbors of each spin is not homogeneous. 
It is distinguished  from a random network due to its scale free behavior,
 $P(k) = k^{-\gamma}$, where $k$ is the number of links of each node and $\gamma$ the
decay exponent of its link distribution\cite{SFN}. Networks with  large $\gamma$ have
 “very famous” nodes (or hubs), i.e., those having direct links to most
other nodes.  This type of network found many realizations in 
real world, from World Wide Webs, power grids, neural and cellular networks, 
till routers of the internet and citation network of scientists.

In addition, an SFN  is also characterized by a  clustering coefficient, $C$,
defined as the average of local clustering, $C_i$,
\begin{equation}
C_i = \frac{2y_i}{ z_i (z_i-1)}
\end{equation}
where $z_i$ is the total number of nodes linked to the $i$-th site and
$y_i$ is the total number of links connecting those nodes. Both $C_i$
and $C$ lie in the interval [0,1]. For a fully connected
network,  $C=C_i=1$, all  nodes connect to each other. 
The parameter $C$ is  related to the number or the density of
triangles.  Because  spins on the vertices of triangular units are frustrated,
$C$ is also related to  the frustration degree of the network.
More comprehensive review on the ubiquitous of SFNs  can be found for example in 
\cite{Newman}.

Thermal averages of the physical quantities of interest are calculated 
using the Replica Exchange algorithm of MC method\cite{Hukushima}. This algorithm   is
implemented to overcome the slow dynamics due to
the presence of local minima in the energy landscape of the system. Slow dynamics 
is  a phenomenon commonly  found in dealing with SGs, where a random walker 
can easily  get trapped in one particular local minimum.  
It is an extended Metropolis algorithm 
where a system is duplicated into  $K$ replicas. All replicas are
simulated in parallel and each is in equilibrium with a heat bath of an 
inverse temperature.  Given $K$ inverse temperatures, $\beta_1, \beta_2, \cdots, \beta_K$, 
the probability distribution of finding the whole system in a state 
 $\{ X \} =
\{ X_1,X_2, \dots, X_K\}$ 
is given by,
\begin{equation}
 P(\{ X, \beta\}) = \prod_{m=1}^{K} \tilde{P}(X_{m},\beta_{m}),
\end{equation}
with
\begin{equation}
\tilde{P}( X_m, \beta_m) = Z(\beta_{m})^{-1} \exp(-\beta_{m} H(X_{m})),
\label{equil}
\end{equation}
and $Z(\beta_m)$ is the partition function of the $m$-th replica.
We can  define an exchange matrix between  replicas as
 $W(X_m,\beta_m| X_n,\beta_n)$, which is the probability
of switching  configuration $X_m$ at the temperature $\beta_m$
with  configuration $X_n$ at $\beta_n$. 

In order to keep the entire system at equilibrium, by using the detailed 
balance condition
\begin{eqnarray}
P( \ldots,\{ X_m, \beta_m \},\ldots, \{ X_n, \beta_n \},\ldots )\cdot
 W(X_m,\beta_m| X_n,\beta_n) \nonumber \\ = P( \ldots,\{ X_n, \beta_m \},
\ldots, \{ X_m, \beta_n \},\ldots )
\cdot W( X_n,\beta_m | X_m,\beta_n), 
\end{eqnarray}
along with Eq.~(\ref{equil}), we have 
\begin{equation}
\frac{ W( X_m,\beta_m | X_n,\beta_n)}{ W( X_n,\beta_m | X_m,\beta_n)}=\exp(-\Delta),
\end{equation}
where $\Delta= (H(X_{m})-H(X_{n})) (\beta_{n}-\beta_{m})$.
With the above constraint we can choose the matrix coefficients
according to the standard Metropolis method, therefore
\begin{equation}
W(X_m,\beta_m| X_{n},\beta_n)=\left \{  \begin{array}{ccc} 
1 & {\rm if} & \Delta \leq 0, \\ 
\exp(-\Delta) & {\rm if} & {\rm Otherwise}.
\end{array} \right.
\label{trans}
\end{equation}
As the acceptance ratio decays exponentially with $(\beta_n-\beta_m)$,
 the exchange is performed   only to the replicas next to each other, 
i.e.,  $W(X_m,\beta_m| X_{m+1},\beta_{m+1})$. 
The replica exchange method is extremely efficient for simulating
systems such as SGs.  
This method has been widely implemented in many complex systems, including
the AF-SFN Ising SG\cite{Bartolozzi}.
In the next section, we discuss the search for SG behaviour of 
Heisenberg antiferrmagnetic system on SFN.
%
\begin{figure}
\begin{center}
\includegraphics[width=0.4\linewidth]{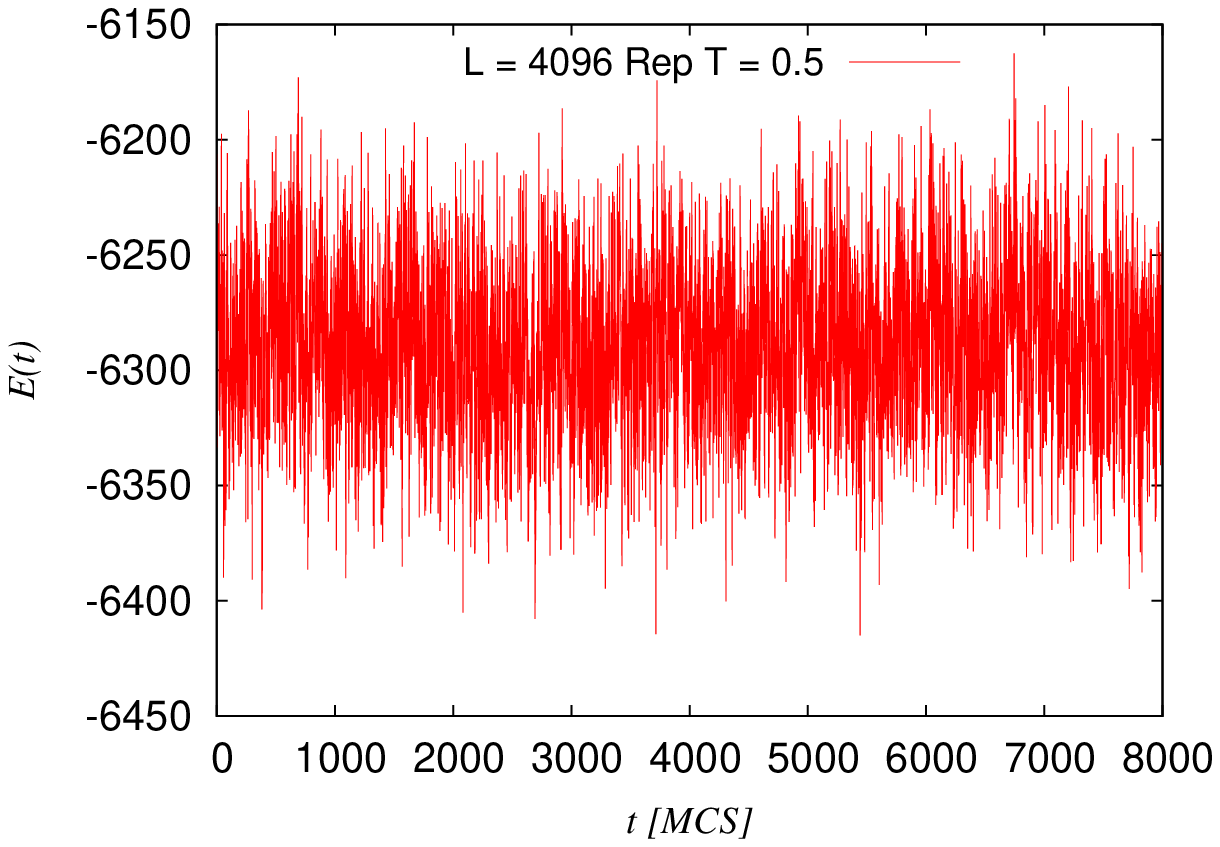}
\includegraphics[width=0.4\linewidth]{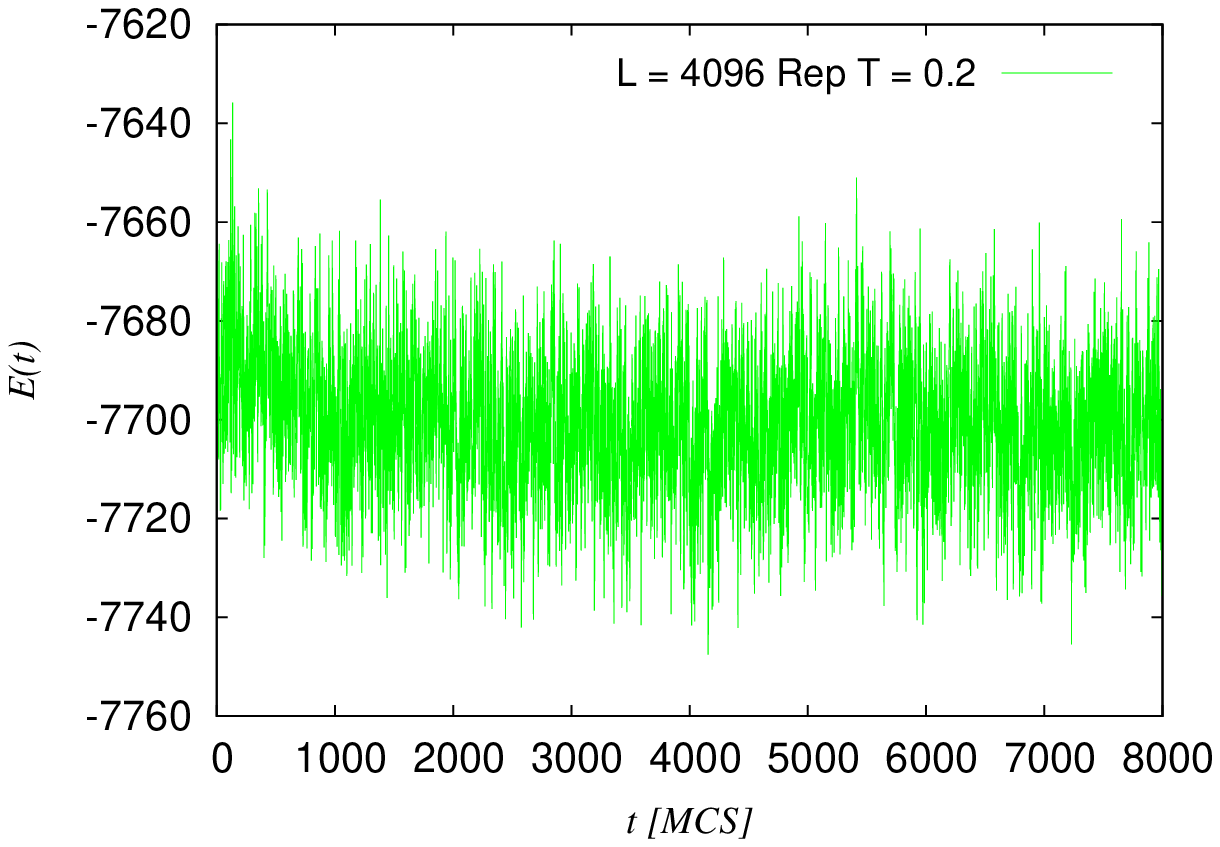}
\caption{Energy time series  of system with $L=4096$ and  $m=5$  at
 $T=0.2$ and $0.5.$}
\label{TS}
\end{center}
\vspace{-0.5cm}
\end{figure}
\vspace{-0.5cm}
\section{Results and Discussion}\label{Results}
\subsection{Equilibrium Behavior}
We have simulated AF Heisenberg model on SFN with various
connectivity densities, $m=2,3, 4$ and $5$.
For each density, we consider several system sizes, $L = 1024, 2048$, and $4096$, which are
 the number of  spins.  Since this is a random system, 
we took many realizations of the networks for each system size, 
then averaged the results over the number of realizations.
Each realization corresponds to one particular connectivity distribution.  
We have to take reasonable number of realizations $N_r$ for the better statistics 
of the results. Previous study on Ising system took $N_r=1000$ realizations\cite{Bartolozzi}. 
Here,  due to the less fluctuation of the  results from 
different realizations, we took moderate number of realizations, i.e., $N_r = 500.$

To check for the  equilibrium behavior of the systems we evaluated the energy time 
series.  In Monte Carlo (MC) simulation, time is assigned as a series of MC steps (MCSs).  
One MCS is defined as visiting each spin once,  either randomly or consecutively, and 
performing a prescribed spin update, i.e.,  Metropolis update.  
We performed $M$ MCSs for each temperature and took $N$ samples out of $M$ MCSs.  
The time series plot of energy for two different temperatures for system size $L = 4096$ 
is shown in Fig. \ref{TS}.  The phenomenon of slow dynamics at lower temperature $(T=0.2)$ 
is shown  in  Fig. \ref{TS} where the average value for the first
half of the plot is larger than that of the rest half. To overcome this problem, 
the equilibration process was increased up to 8000 MCSs.

\begin{figure}[h]
\begin{center}
\includegraphics[width=7.8cm,height=5cm]{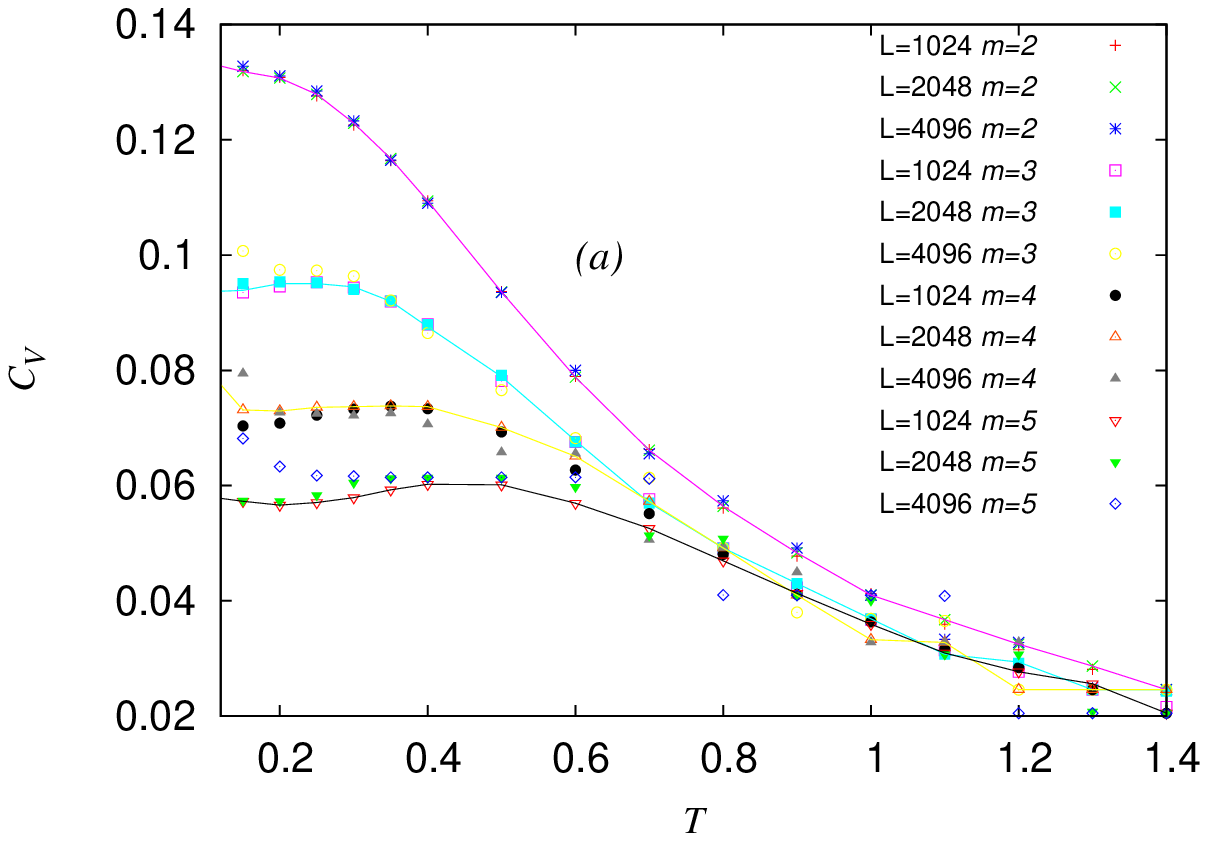}
\includegraphics[width=7.8cm,height=5cm]{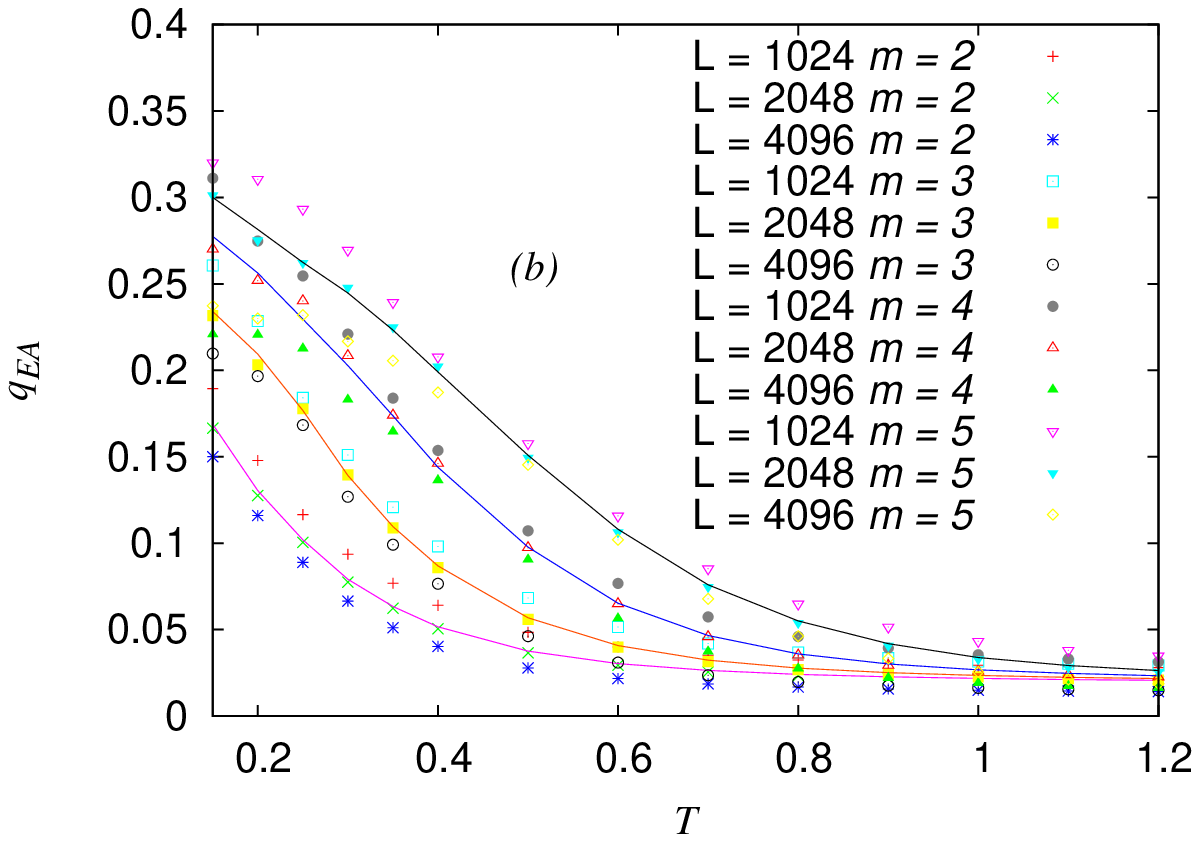}
\vspace{-0.5cm}
\caption{Temperature dependence of (a) specific heat and (b)
SG order parameter for different sizes
with $m=$2,3,4 and 5. The solid lines are guides to the eye 
for the clarity  of peaks.}
\label{sp_heat}
\end{center}
\vspace{-0.6cm}
\end{figure}
We  also calculated the specific heat which  is defined as follows
\begin{equation}
C_v = \frac{N}{kT^2} \left( \langle E^2 \rangle - \langle E \rangle^2\right)
\end{equation}
where $L$, $k$ and $\langle E \rangle$ are respectively the number of nodes, 
Boltzmann constant and the ensemble average of energy.
The temperature dependence of the specific heat for various system sizes
is shown in Fig.~\ref{sp_heat}(a), where the statistical errors are comparable 
to the size of the symbols.  Although the specific heat plot exhibits  no singularity,
there is a maximum value at intermediate  temperatures.  
As indicated, the peaks of the specific heat shift to higher temperatures as the 
increase in $m$.  The presence of peak may signify the existence 
of phase transition. A clear sign of the SG phase can be observed from the plot 
of overlapping order parameter which will be presented in the next subsection.

\subsection{Spin Glass Order Parameter}
To search for the SG phase transition, we calculate SG order parameter defined as follows
\begin{equation}
q_{EA} =  \langle \left|\sum_i \vec s_i^{\alpha} \otimes 
\vec s_i^{\beta} \right| \rangle_{av}
\end{equation}
where $\vec s_i$ is the spin on node $i$-th while  $\alpha$ and $\beta$  denote two sets of replicas.
The origin of this quantity is  from the scalar product of the vector spins.
A scalar product of two vectors will be maximum (zero) if they are parallel
(perpendicular). This idea is implemented to capture the frozeness of 
the spin configuration.   
If a system is frozen, the spin configurations of two replicas with
the same inverse temperature from different sets will be more or less
similar, regardless of globally rotational invariant.  As explained in Sec. \ref{Model}, 
system is replicated into K replicas, each belongs to inverse temperature. 
For the sake of $q_{EA}$ calculation,  the whole replicas are duplicated, therefore
we have two sets of replicas. Each set contains $K$ replicas which are
exchanged during the simulation. Only replicas from the same set were  
exchanged during the simulation.  

The overlapping parameter will give finite value if system in SG phase.
This  applies for any SG model, including the  Ising and the Heisenberg model.
For Ising model, $q_{EA}$ is simply the multiplication of the overlapped spins.  
In contrast for the Heisenberg model where 
 spins are allowed to rotate in any direction, we take the tensor product
instead of the scalar product, resulting $q_{EA}$ with nine components.
The plot of temperature dependence of $q_{EA}$ for various 
different system sizes is shown in Fig. \ref{sp_heat}(b). As indicated,
 this parameter increases as temperature decreases, which is the evidence for the existence of
 SG phase at lower temperatures.  
\begin{figure}
\begin{center}
\includegraphics[width=7.8cm,height=5cm]{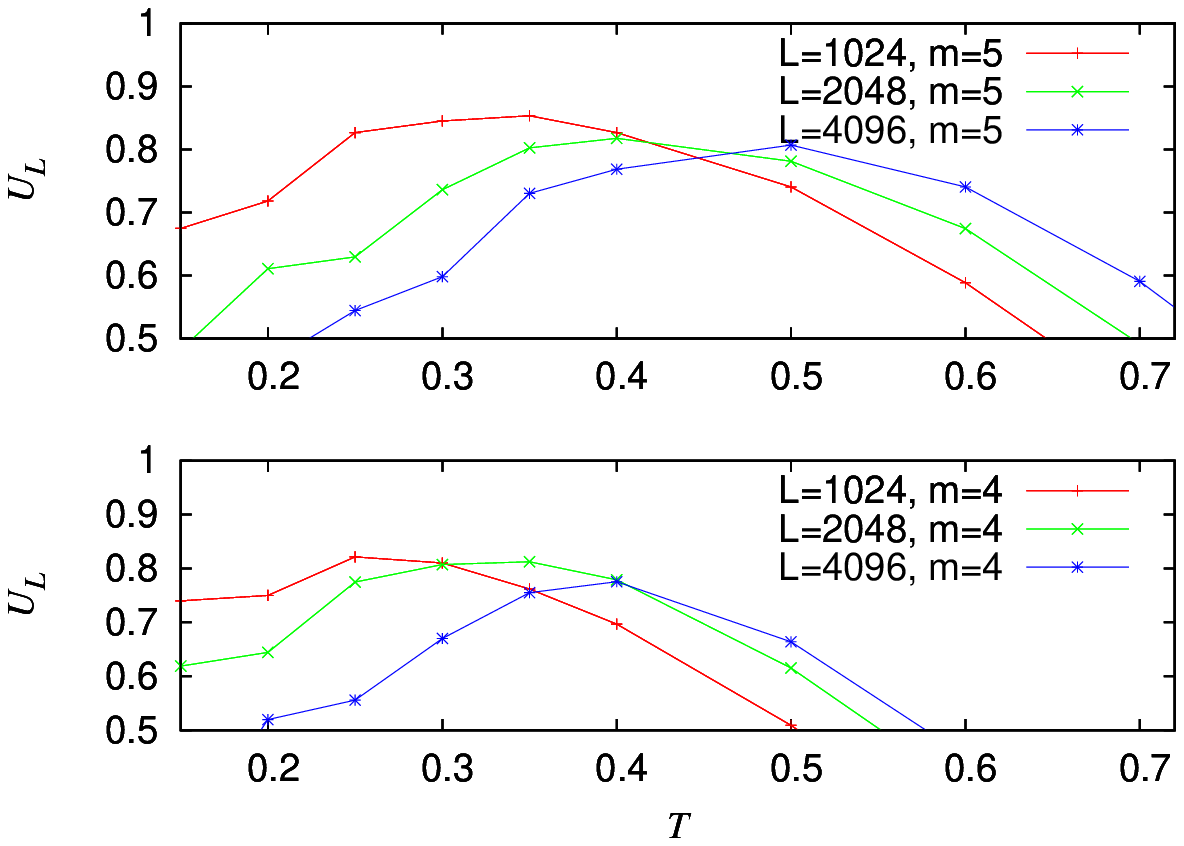}
\includegraphics[width=7.8cm,height=5cm]{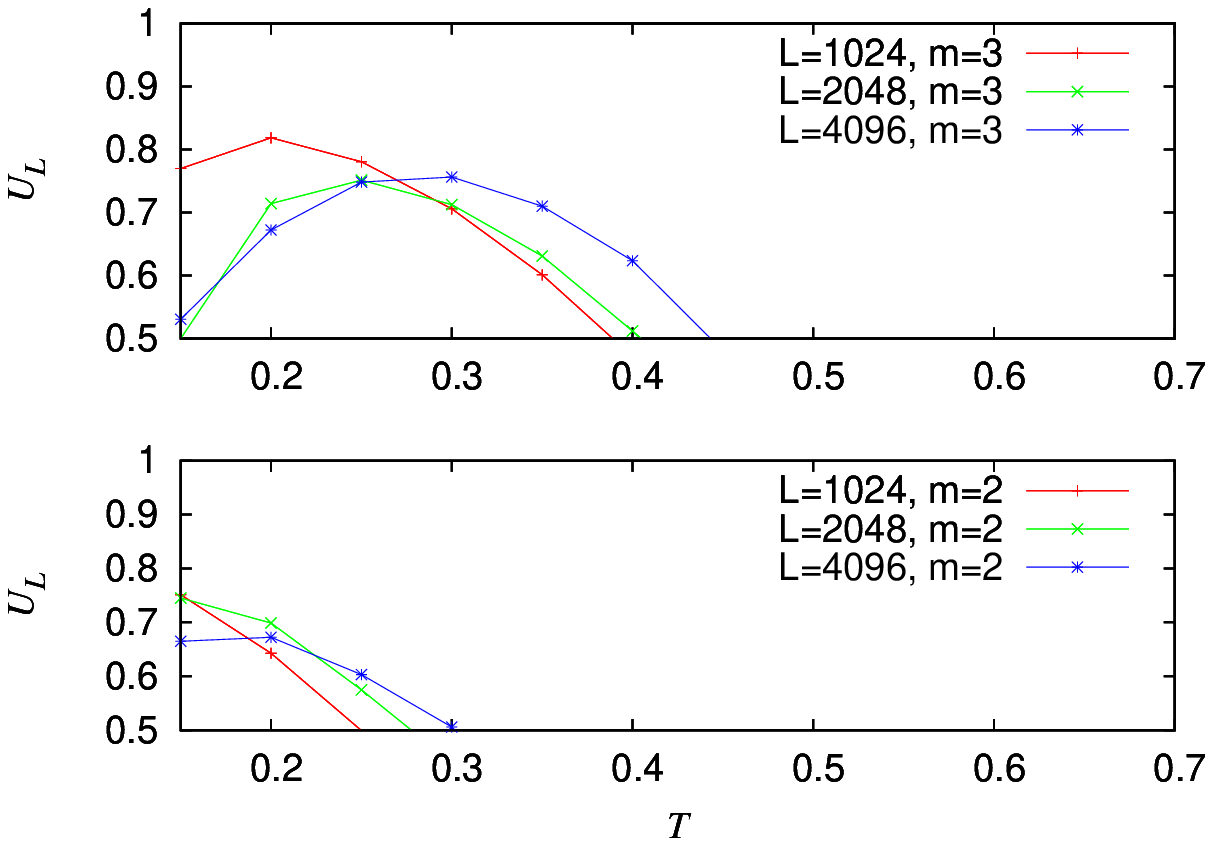}
\vspace{-0.5cm}
\caption{Binder parameter of  $q_{EA}$ for various values
of $m$. Solid lines are guide to the eyes.}
\label{Binder}
\end{center}
\vspace{-0.5cm}
\end{figure}
\vspace{-0.2cm}

To clarify that this a true SG phase, we calculate
the cumulant ratio (Binder parameter) \cite{Binder} of  $q_{EA}$ defined as follows
\begin{equation}
U_L = \frac{1}{2}\left[ 11 -\frac{9\langle q^4\rangle}{\langle q^2 \rangle^2} \right]
\end{equation}
The plot of $U_L$ for different system sizes is  shown in Fig. \ref{Binder}.  
It is clear that there is a single crossing point, emphasizing the existence of SG phase transition.
However, due to the anomaly of the crossing pattern, which is different from the standard ones,
such as in the previous study on Ising model\cite{Bartolozzi}, we did 
not perform scaling plot of $U_L$ for the estimate of critical temperature and exponent.
Instead, we  roughly estimate the
critical temperature for each connectivity density as listed 
in Table 1, where numbers in bracket are the uncertainty for the last
digits. There is a systematic increase of critical temperature
as the connectivity density increases. This is consistent with
the Mean Field theory argument, where an ordered phase for a system with large 
connectivity tend to be more stable.
\section{Summary and Conclusion}\label{Summary}
In summary, we have studied
AF Heisenberg  model on scale free network, using Replica Exchange MC method.
We simulated several different connectivity densities ($ m=2,3,4$ and $5$)  
and calculated such physical quantities as ensemble average of energy, the specific 
heat, the overlapping parameter and its  cumulant ratio (Binder parameter).
A sign  for  finite  SG phase transition  was observed for
all values of $m$. There is a systematic increase of $T_c$ as $m$ increases.
This is related to the fact that systems with large connectivity
tend to be more robust against thermal fluctuation. The existence of SG phase 
even in the system with $m = 2$ is generally in a good agreement with 
the case for canonical systems where Heisenberg SG phase was clearly
observed in system with large dimension, e.g. the 4D system\cite{Coluzzi}.
It is therefore interesting to search for  the lower critical value of $m$
by studying systems with fractional value  of $m$.  
We will consider such system in our future study.
\begin{table}
\vspace{-0.5cm}
\caption{The estimate of critical temperature $T_c$
 for each value of $m$.$\\$}
\label{Table01}
\begin{center}
\begin{tabular}{c|c}
\hline
\hline
Connectivity density & $T_c$  \\
($m$) &    \\
\hline
5 & $0.44(3)$\\
4 & $0.35(5)$\\
3 & $0.24(6)$\\
2 & $0.19(6)$\\
\hline
\end{tabular}
\end{center}
\vspace{-0.5cm}
\end{table}
\vspace{-0.2cm}

\section*{Acknowledgments}

The authors wish to thank K. Hukushima, M. Troyer and Y. Okabe for valuable discussions.
The computation of this work was performed
using parallel computing facility in the Department of Physics
Hasanuddin University and the HPC facility of the Indonesian
Institute of Science. The work is supported by the  HIKOM
research grant 2014 of the Indonesian Ministry of Education and Culture.


\section*{References}
\begin{thebibliography}{9}
\bibitem{Cannella} V. Cannella and J. A. Mydosh, Phys. Rev. B, {\bf 6},4220 (1972)
\bibitem{SK}D. Sherrington and S. Kirkpatrick,
Phys. Rev. Lett. {\bf 35}, 1792 (1975).
\bibitem{Edwards}S. F. Edwards and P. W. Anderson, J. Phys. F 5, 965, (1975).
\bibitem{Thirumalai}T. R. Kirkpatrick and D. Thirumalai,
Phys. Rev. Lett. {\bf 58}, 2091, (1987).
\bibitem{Kawamura} H. Kawamura and S. Nishikawa, Phys. Rev. B {\bf 85}, 134439 (2012).
\bibitem{Wittmann} M. Wittmann and A. P. Young, Phys. Rev. E {\bf 85}, 041104, (2012)
\bibitem{Nishimori}H. Nishimori, {\it Statistical Physics of Spin Glasses and Information
Processing}, Oxford Univ. Press, (2001).
\bibitem{Villain} J. Villain, J. Phys. C  {\bf 10}, 1717 and 4793 (1977).
\bibitem{Tasrief2004} T. Surungan, Y. Okabe, and Y. Tomita, J. Phys. A {\bf 37}, 4219, (2004).
\bibitem{Bartolozzi} M. Bartolozzi, T. Surungan, D.B. Leinweber and
   A.G. Williams,  Phys. Rev. B{\bf 73}, 224419 (2006).
\bibitem{Herrero}C. P. Herrero, Phys. Rev. E {\bf 77}, 041102 (2008).
\bibitem{SFN}A.-L. Barabasi and R. Albert, Science 286, 509 (1999).
\bibitem{Newman} M. Newman, A-L. Barabasi and D. J.  Watts, {\it The Structure and Dynamics 
of Networks}, Princeton Uni. Press, (2006).
\bibitem{Coluzzi}B. Coluzzi, J. Phys. A: Math. Gen. 28 , 747 (1995).
\bibitem{Matsubara}F. Matsubara1, T. Shirakura, S. Endoh and S. Takahashi,
 J. Phys. A: Math. Gen. 36 10881, (2003).
\bibitem{Hukushima}K. Hukushima and K. Nemoto, J. Phys. Soc. Jpn. {\bf 65}, 1863, (1996).
\bibitem{Kawashima}N. Kawashima and A. P. Young, Phys. Rev. B {\bf 53}, R484, (1996).
\bibitem{Binder} K. Binder, Z. Phys. B{\bf 43}, 119 (1981);
Phys. Rev. Lett. {\bf 47}, 693 (1981).
\end{thebibliography}
\end{document}